\documentclass[utf8]{frontiersFPHY} 

\usepackage{url,hyperref,lineno,microtype,subcaption,tikz}
\usepackage{mathtools}
\usepackage[onehalfspacing]{setspace}

\def\keyFont{\fontsize{8}{11}\helveticabold }
\def\firstAuthorLast{Jamieson {et~al.}} 
\def\Authors{Blair Jamieson\,$^{1,*}$, Matt Stubbs\,$^{1}$, Sheela Ramanna\,$^{1}$, John Walker\,$^{1,2}$, Nick Prouse\,$^{2}$, Ryosuke Akutsu\,$^{2}$, Patrick de Perio\,$^{3}$ and Wojciech Fedorko\,$^{2}$}
\def\Address{$^{1}$University of Winnipeg, Winnipeg, Manitoba, Canada,
$^{2}$TRIUMF, Vancouver, British Columbia, Canada,
$^{3}$Kavli IPMU (WPI), UTIAS, The University of Tokyo, Tokyo, Japan}
\def\corrAuthor{Blair Jamieson}

\def\corrEmail{bl.jamieson@uwinnipeg.ca}

\begin{document}
\onecolumn
\firstpage{1}

\title[Machine Learning to improve neutron ID in water Cherenkov detectors]{Using Machine Learning to Improve Neutron Identification in
Water Cherenkov Detectors} 

\author[\firstAuthorLast ]{\Authors} 
\address{} 
\correspondance{} 

\extraAuth{}

\maketitle

\begin{abstract}

\section{}

Water Cherenkov detectors like Super-Kamiokande, and the next generation Hyper-Kamiokande are adding gadolinium to their water to improve the detection of neutrons.  By detecting neutrons in addition to the leptons in neutrino interactions, an improved separation between neutrino and anti-neutrinos, and reduced backgrounds for proton decay searches can be expected.  The neutron signal itself is still small and can be confused with muon spallation and other background sources. In this paper, machine learning techniques are employed to optimize the neutron capture detection capability in the new intermediate water Cherenkov detector (IWCD) for Hyper-K. In particular, boosted decision tree (XGBoost), graph convolutional network (GCN), and dynamic graph convolutional neural network (DGCNN) models are developed and benchmarked against a statistical likelihood-based approach, achieving up to a 10\% increase in classification accuracy. Characteristic features are also engineered from the datasets and analyzed using SHAP (SHapley Additive exPlanations) to provide insight into the pivotal factors influencing event type outcomes. The dataset used in this research consisted of roughly 1.6 million simulated particle gun events, divided nearly evenly between neutron capture and a background electron source.

\tiny
 \keyFont{ \section{Keywords:} machine learning, graph neural networks, water Cherenkov detector, particle physics, neutrino physics} 
\end{abstract}

\section{Introduction}

One exciting frontier within experimental neutrino physics is the improved identification of neutrons from inverse beta decay reactions ($\nu_{e} + p^{+} \rightarrow e^{+} + n$). This task, referred to as `neutron tagging,' is particularly challenging due to the low energy scale and faint signals involved. Progress in this field could lead to a host of advancements in particle physics, including a first detection of diffuse supernova background neutrinos \citep{FERNANDEZ2016353}, and improved understanding of the neutrino mass hierarchy and the CP violating phase \citep{irvine}. However, water Cherenkov (WC) detectors have historically been limited in their detection capability of these low energy neutron capture events. 

Neutrons are commonly liberated in water due to the inverse beta decay (IBD) process, in which an electron antineutrino collides with a proton to yield a positron and a free neutron. From there, the free neutron undergoes thermalization, colliding with neighbouring molecules and gradually losing energy until it reaches water temperature. Approximately 200 $\mu$s after thermalization, the free neutron is captured by a proton or oxygen nucleus, releasing a gamma particle $\gamma$ at 2.2 MeV ($n + p \rightarrow d + \gamma$) \citep{WATANABE2009320} where $d$ is deuterium (or `heavy hydrogen'), the isotope of hydrogen with a proton and neutron in the nucleus. The capture cross-section of this neutron capture on a hydrogen nucleus (proton) is only 0.33 barns, and the resulting 2.2 MeV gamma produces such a faint light signal that it is very difficult to identify by a PMT in a WC detector. Many traditional WC detectors actually have thresholds of 5 MeV, high enough that none of these signals would be recorded at all.

To address this problem, the addition of gadolinium chloride (GdCl$_{3}$) (a light, water soluble-compound) to the SK detector water was proposed in 2003 \citep{PhysRevLett.93.171101}. Gadolinium is known for having the ``largest capture cross-section for thermal neutrons among all stable elements'' \citep{PhysRevLett.108.052505}. At approximately 49700 barns, the gadolinium capture cross-section is over six orders of magnitude larger than for free protons, leading to faster captures. Neutron capture on gadolinium also leads to an 8 MeV cascade of gammas (7.9 MeV cascade 80.5\% of the time and an 8.5 MeV cascade 19.3\% of the time \citep{WATANABE2009320}, a signal which is far easier to detect due to its relatively higher energy. Beacom and Vagins showed that only a 0.1\% addition of gadolinium by mass leads to at least a 90\% probability of neutron capture on gadolinium (the other 10\% or less of neutron captures are still by hydrogen nuclei). In addition, the neutron capture by gadolinium after thermalization occurs in roughly 20 $\mu$s, nearly 10 times more quickly than capture on protons. 

This paper presents the implementation of several machine learning methods that attempt to improve the efficiency of neutron tagging for simulations of neutron capture and background radiative neutrino events within the gadolinium-doped intermediate WC detector (IWCD) for Hyper-K \citep{protocollaboration2018hyperkamiokande}. The structure of the paper is as follows. Section 2 discusses related works in the intersecting fields of particle physics, neutron tagging and machine learning. Section 3 then introduces the relevant machine learning theories and algorithms used in this research, including boosted decision trees (XGBoost), SHAP (SHapley Additive exPlanations) and graph neural networks (GNNs). In Section 4, the data and data simulation process are explored. Also in this section, a likelihood analysis benchmark is shown based on event hit totals and charge sums. Section 5 illustrates the process of engineering characteristic features from the data and covers the implementation and tuning of the XGBoost model. Afterward, an analysis of relative feature importances is applied using SHAP. Section 6 presents the results of the  GCN and DGCNN graph neural network models and discusses various methods of graph network construction.  One of the main goals of this research is to investigate the applicability, performance and feasibility of GNNs on the IWCD particle data, in particular for the low energy regime where the number of event hits is small and CNNs tend to struggle. Finally, Section 7 concludes on the findings of the previous chapters.

\section{Related work}

\subsection{Machine Learning in Particle Physics}

The uses of machine learning and its historical development in the field of particle physics is discussed in \citet{bourilkov}. Traditional means of event selection in particle physics are discussed in both \citet{bourilkov} and \citet{guest}. These methods often involved a series of boolean `cuts' (decisions) on single variables at a time, followed by statistical analyses on the remaining data. However, over the past several decades, physicists have developed algorithms that employ machine learning to study multiple variables simultaneously in multivariate analysis (MVA). \citet{guest} describes the use of an assortment of machine learning techniques for MVA in the physics context, include support vector machines, kernel density estimation, random forests, boosted decision trees, etc. \citet{carleo} provides an overview of applications of machine learning within the physical sciences, including applications to quantum computing, chemistry and cosmology. \citet{carleo} also discusses applications to particle physics, including jet physics and neutrino signal classification. Machine learning applications are discussed for a variety of neutrino experiments, including the MicroBooNE collaboration, Deep Underground Neutrino Experiment (DUNE) and the IceCube Observatory at the South Pole.

\subsection{Boosted Decision Trees}

Boosted decision trees (BDTs) are a commonly applied machine learning method in modern particle physics analysis. For example, \citet{bdt1} details the improved performance of particle classification in the MiniBooNE experiment, which searches for neutrino oscillations, using BDTs compared to artificial neural networks. \citet{radovic} discusses multiple use cases of BDTs at the Large Hadron Collider (LHC) at CERN, including the application of BDTs to improve the energy reconstruction (mass resolution) of the CMS (Compact Muon Solenoid) calorimeter, as well as the implementation of BDTs to improve the sensitivity of the ATLAS detector to various Higgs boson decay modes. For the latter, the sensitivity of diphoton decay ($H$ $\rightarrow$ $\gamma$$\gamma$) and antitau-tau pair decay ($H$ $\rightarrow$ $\tau^{+}$$\tau^{-}$) were improved by an amount equivalent to adding 50\% and 85\% more data to the detector, respectively. Beyond learning tasks, BDTs can also be used at the early stages of the machine learning lifecycle. For example, \citet{bdt2} modifies the standard boosted decision tree algorithm to improve high-level triggering in detector data acquisition systems. A general BDT usage guidebook is presented in \citet{bdt3} for the hypothetical identification of the smuon particle and performance is compared to the classic `cut-and-count' approach.

\subsection{Deep Learning and Graph Neural Networks}

The computer vision approach to particle classification, which consists of reconstructing particle events as images and applying convolutional neural networks (CNNs), has been applied in various detector experiments \citep{andrews, macaluso, brickwedde}. However, the conversion of data from irregular detector geometries into a 2-dimensional grid for images inherently causes loss of information. For events with few hits, the sparsity of the resulting image is also difficult for CNNs to learn from. Alternately, deep learning sequence models, inspired by tasks in natural language processing, have also been adapted to the particle physics domain by modelling particles and measurement objects in a sequential order. Instances of this approach include tagging of jets containing $b$-hadrons in the ATLAS experiment \citep{atlasrnn} and classifying energetic hadronic decays in the CMS experiment \citep{cmsdecay}. However, the imposed ordering of objects in the sequence constrains the learning of the model. The limitations of both computer vision and sequence deep learning approaches are discussed in \citet{shlomi}.

Graph neural networks (GNNs) represent an emerging architectural class of deep learning which undertakes to learn from data structured in a graph format, for which particle events find a natural representation. \citet{shlomi} surveys the theory and applications of GNNs in particle physics. The graph classification task is partitioned into jet classification and event classification. While jets represent a part of a particle collision occurrence, an event references the full history of the particular physics process. In \citet{qujets}, the jet is viewed as an unordered structure of particles, analogous to the point cloud representation of shapes in 3D space. The authors propose the `ParticleNet’ method, which uses the `EdgeConv' block as an analogue for CNN convolution on 3D point clouds and updates the graph representation dynamically, and report state-of-the-art performance on jet tagging tasks. For event classification, one example is the deployment of GNNs in the IceCube neutrino observatory \citep{choma}. In this case, the irregular hexagonal geometry of the detector is itself modeled as a graph, where the sensors are the graph nodes and the edges represent their connections. Given the sparsity of activated sensors in an event, every event is considered as a different graph composed only of the active sensors in the event. Although learning occurs over relatively small sample sizes, the authors report an approximate 3x improvement in signal-to-noise ratio compared to the physics baseline and the CNN approach.

\section{Machine Learning Methods Studied}

\subsection{XGBoost}
\label{decision trees}

Over the last several years, the machine learning model `XGBoost' has gained popularity for its performance in classification or regression tasks involving tabular data over a variety of domains, including vehicle accident detection \citep{xgboost-highways}, cancer diagnostics \citep{cancer}, network intrusion detection \citep{intrusion-detection} and Higgs boson identification \citep{higgs}. XGBoost stands for `eXtreme Gradient Boosting'. In general, gradient boosting refers to the process of beginning with a single weak learner and iteratively constructing superior learners that improve on the errors of their predecessors. The new learners attempt to optimize an overall loss function over the problem space by each following the negative gradient of the loss function.

XGBoost was introduced by Chen and Guestrin in their 2016 paper  \citep{xgboost-original}, which considered the case of decision trees as the individual learners in the function ensemble. In general, a decision tree applies classification or regression to an example by partitioning the example through a series of splits (decisions) from the root node to a leaf of the tree. The given tree splits are themselves computed by calculating which partition leads to maximum information gain. For any specific training example, the overall output is the additive sum of the outputs from every individual tree. To apply gradient boosting in the context of decision trees, an appropriate objective function (loss) must be defined. Chen and Guestrin define the overall objective function as the sum of a regular loss and a regularization term.  Practically, when constructing a given decision tree in the XGBoost ensemble, it is too computationally expensive to iterate through all possible tree structures and compute the objective function for each possibility. Instead, a greedy approach is applied where, starting at the tree node, branches are successively added by finding the particular split which leads to maximum gain. 
\subsection{SHAP}
\label{Shapley values}

The Shapley value (which SHAP derives from) traces back to Lloyd Shapley's paper ``stochastic games," published in Princeton in 1953 \citep{shapley-original}. At the time, Shapley was studying the field of cooperative game theory and searching for a mapping from a coalition single game to a numeric payoff vector. Shapley found an intuitive solution to the seemingly intractable problem by searching for a set of ``reasonable axioms" (efficiency, symmetry, dummy and additivity) \citep{shapley-original}. His resulting `Shapley value' can be viewed as an ``index for measuring the power of players in a game" \citep{shapley-explanation}. In the context of physical event classification, the player is analogous to the event feature, the game is analogous to the event and the label is the analogous to the numeric payoff output.

Winter's paper \citep{shapley-explanation} reviews the theoretical framework for the derivation of the Shapley values. Lundberg and Lee \citep{SHAP} extend this definition, introducing the `SHAP' values as the Shapley values of a ``conditional expectation function of the original model." They also present the concept of the `explanation model' in which the output prediction of the ML model may be viewed as a model itself. Their definition of an `\textit{Additive Feature Attribution Method}' is one in which the explanation model may be represented as a linear function of binary variables. This makes it possible to view the marginal contributions of individual features for any given event.

\subsection{Graph Neural Network (GNN)}

While traditional machine learning algorithms have proven effective at learning from tabular data, they have historically struggled to learn well from natural data, including images, natural language or audio. While deep learning architectures like convolutional neural networks (CNN) \citep{convnet} and recurrent neural networks (RNN) \citep{RNN} have proven effective at learning from image or sequence data, \textit{geometric deep learning}, the umbrella term for the task of deep learning on graph data, is an emerging area of research. Where a given graph G may be denoted by its set of vertices and edges $G = \{V, E\}$, the nodes represent objects or concepts and the edges represent their relationships. A variety of situations may be modelled by graphs, including social networks, molecules, Internet traffic, etc \citep{gnnreview}. The GNN is designed to operate directly on data input as a graph. Low energy neutrino-induced events in the IWCD may be naturally represented by a graph, where the PMTs constitute the nodes and the edges represent the connections between the PMTs.

The origin of deep learning on graphs traces back to the late 1990s, when RNNs were applied to directed, acyclic graphs (directional edges, no loops formed by a collection of edges) \citep{gnnreview}. Using this approach, node feature states are updated in successive layers until equilibrium is reached. This technique was later generalized to cyclic graphs as well in 2008 \citep{scarselli}. Soon after, following the widespread success of CNNs, significant interest grew in generalizing some concepts from CNNs to learning on graphs. The first successful adaption of the convolution operation to graphs was developed by Bruna et al.\ in 2013 using Laplacian eigenvectors (\cite{bruna}). The computational complexity of this procedure was later greatly reduced by applying polynomial spectral filters instead of Laplacian eigenvectors (\cite{defferard, kipf}). Approaches have also been developed which apply spatial, and not spectral, filters for the convolutional operation (\cite{masci}). In general, GNNs apply a series of filtering and activation layers to update the feature representation of every node. Once the network has passed all the hidden layers, the output node labels may be used directly in node-focused tasks, or the node outputs may be pooled together to obtain an overall coarsened representation for graph classification.

\subsubsection{Graph Convolutional Network}
\label{gcn description}

Kipf and Welling demonstrated the successful approach of using a convolutional architecture to learn on graphs in their paper ``Semi-supervised classification with graph convolutional networks" \citep{kipf}. This approach applies an approximation of spectral graph convolution. The spectral decomposition of a graph denotes the breakdown of the graph's Laplacian matrix $\mathcal{L}$ into its elementary orthogonal components, i.e. the \textit{eigendecomposition} of $\mathcal{L}$. The graph Laplacian $\mathcal{L}$ represents a graph in matrix format and is a graphical analogue to the familiar Laplacian operator for multivariate and continuous functions. For a graph $G = \{V, E\}$, $\mathcal{L}$($G$) is equal to the difference between the degree matrix $D$ (diagonal matrix where every element represents the degree, i.e. number of connections of the corresponding vertex) and adjacency matrix $A$ (matrix with vertices labelled by rows and columns where 0s and 1s represent nonadjacent and adjacent pairs of vertices) of $G$. However, the computation of  $\mathcal{L}$ is computationally expensive and can be a procedural bottleneck.  \citet{Hammond} proposed a computation of $\mathcal{L}$ using the first K Chebyshev polynomials that avoids diagonalization. By taking the first-order Chebyshev approximation K=1 and further constraining other parameters, the multi-layer GCN propagation rule is reached,


\begin{equation}
\label{prule}
    H^{l+1} = \sigma(\tilde{D}^{-\frac{1}{2}}\tilde{A}\tilde{D}^{-\frac{1}{2}} \; H^{l}\; W^{l}),
\end{equation}

\noindent where $H^l$ and $H^{l+1}$ denote the node feature matrices at layers $l$ and $l+1$, $\tilde{A}$ = $A$ (adjacency matrix of graph) + $I_{N}$ (identity matrix), $\tilde{D}_{ii}$ = $\sum{_j}\tilde{A}_{ij}$, $W^{l}$ denotes the matrix of weights at layer $l$ and $\sigma$ is an activation function such as the rectified linear activation unit (ReLU).

\subsubsection{Dynamic Graph Convolutional Neural Network}
\label{dgcnn section}

The dynamic graph convolution neural network (DGCNN), introduced by Wang et al. \citep{dgcnn}, was designed specifically to learn from point cloud graphs for segmentation or classification tasks. Point clouds are collections of three-dimensional coordinates (points) in Euclidean space. However, the DGCNN model also allows the graph nodes to include other features in addition to the spatial coordinates. The main feature of the DGCNN model is the introduction of the `EdgeConv' convolutional operator. EdgeConv is designed to learn edge features between node pairs, i.e. a node and its neighbouring connections. The DGCNN model is dynamic because, for every EdgeConv block, the graph representation is updated. This departs from the action of operating on a fixed graph like most other GNN architectures.

In the DGCNN model, a series of EdgeConv layers are applied to the graph. For a given layer in the network, the EdgeConv operation is applied for every node and its $k$ nearest neighbours in semantic space, where $k$ is a tunable hyperparameter. For two neighbouring nodes $\mathbf{x}_{i}$ and $\mathbf{x}_{j}$, a fully connected layer $h_{\Theta}()$ with learnable weights $\Theta$ and an adjustable number of compute units is applied to learn the pairwise edge features $\mathbf{e}_{ij}$. The node representations are then updated by aggregating these edge features over the node neighbourhood. The EdgeConv filter $h_{\Theta}(\mathbf{x}_{i}, \mathbf{x}_{j}) = h(\mathbf{x}_{i}, \mathbf{x}_{j} - \mathbf{x}_{i})$ operates over individual nodes and local node neighbourhoods, thus allowing the model to learn both local neighbourhood structure and global graph structure. In addition, the dynamic recomputation of the graph for every EdgeConv layer allows for groupings of nodes in semantic space compared to the fixed spatial input space, allowing for a diffusion of information throughout the entire graph.

\section{Data Analysis}
\subsection{Data Simulation}
\label{data simulation}

The data used in this research was simulated using WCSim software to generate neutron and background electron events for the IWCD detector. WCSim, designed to recreate physics events within large WC detectors \citep{WCSim}, is based on Geant4 \citep{Geant4} and also depends on ROOT \citep{ROOT}. The simulations used a cylindrical tank with a height of 6 m and a diameter of 8 m, and with 525 multi-PMT (mPMT) modules of 19 Hamamatsu PMTs each lining the walls of the simulated detector. With a PMT dark noise rate of 1~kHz and gadolinium doping of 0.1\% by mass in the water to generate an approximate 90\% thermal neutron capture on gadolinium nuclei, the simulations procured datasets of about 1.6 million events in total divided nearly evenly between neutron capture and background electron events. The data was saved in a 3-dimensional format of (event, hit, features). where the eight feature values stored were the charge, time, 3D position (x, y, z) and 3D orientation (dx, dy, dz) of each hit PMT. Other features may be engineered from these base eight, a topic which is explored in Section \ref{feature engy}.

Multiple datasets were tested with electron background distributions of different uniform energy levels (i.e. a uniform 8 MeV or 20 MeV energy distribution).  Comparisons with the different background distributions can be found in the thesis of~\cite{matt}. However, to generate a more realistic approximation of background in the IWCD, an electron background was simulated according to an approximate muon spallation source.  Only this more realistic background is used in this paper. In her presentation on muon spallation background in the Super-Kamiokande experiment, Bernard notes that at lower energy scales (tens of MeVs), muon spallation is the dominant source of background \citep{lauraspallation}. Due to the high muon flux at sea level of $6.0 \ast 10^{5}$ $\textrm{m}^{-2}$ $\textrm{hr}^{-1}$ \citep{muon2}, SK was built under 1000 metres of rock. The muons lose energy as they travel through the rock, leading to a far reduced flux rate of $9.6$ $\textrm{m}^{-2}$ $\textrm{hr}^{-1}$ at the detector. The IWCD, however, is to be deployed in only a 50 m deep pit. Therefore, the spallation flux will be greater for IWCD, and it is even more important to reduce this background for identifying neutron captures at low energies. The combined muon spallation energy spectra from Bernard was used as an input to WCSim, replicating the SK spallation energy distribution for the simulation of electron background radiation events in the IWCD detector. The resulting electron background energy distribution follows a right-skewed distribution from approximately 0 to 16 MeV. This background, along with the regular neutron capture events generated by WCSim, constituted the dataset used in this research.

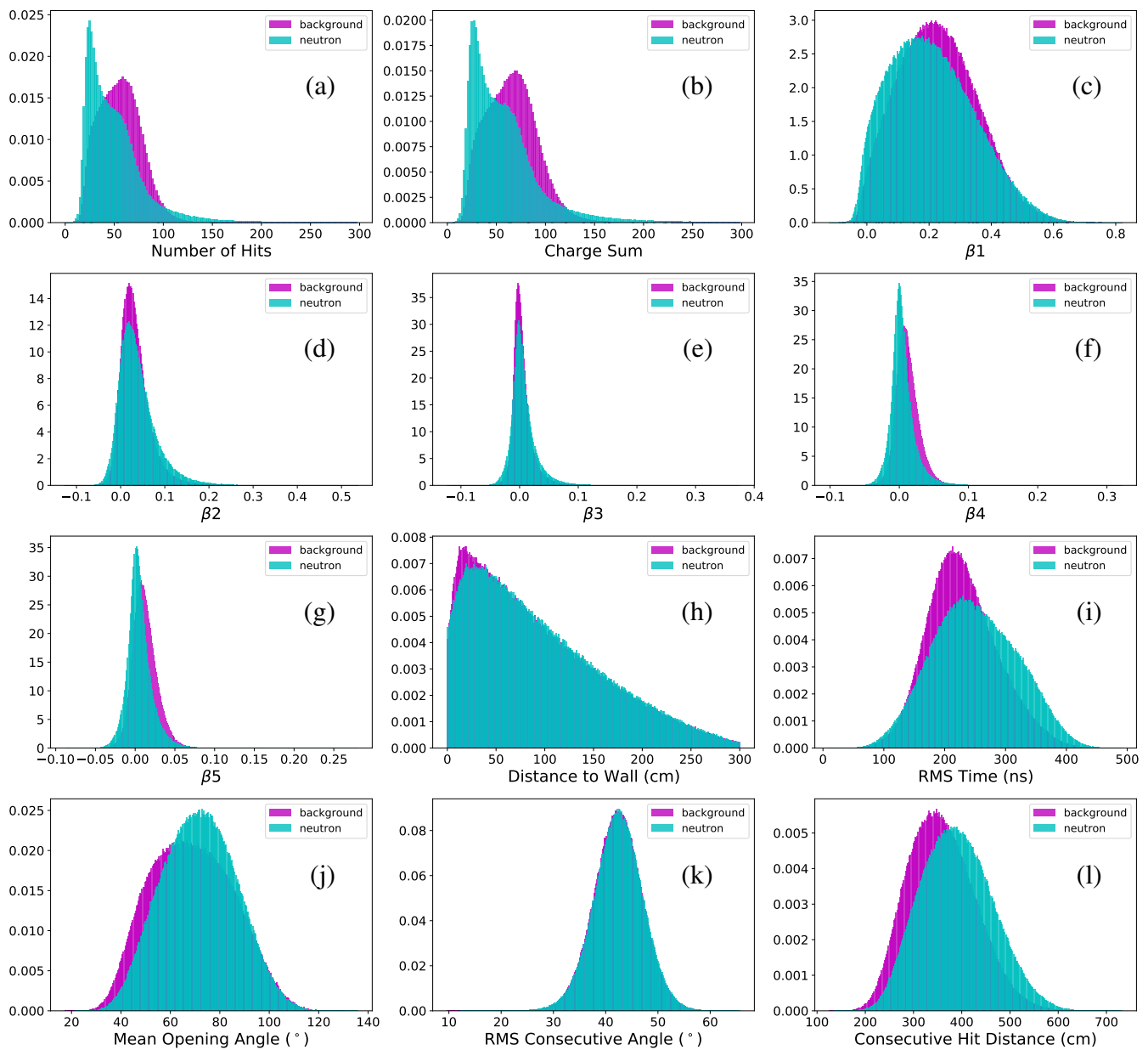
\begin{figure}[htb!]
    \centering
\begin{tikzpicture}
 \node[anchor=south west,inner sep=0] (image) at (0,0) { \includegraphics[width=1.0\textwidth]{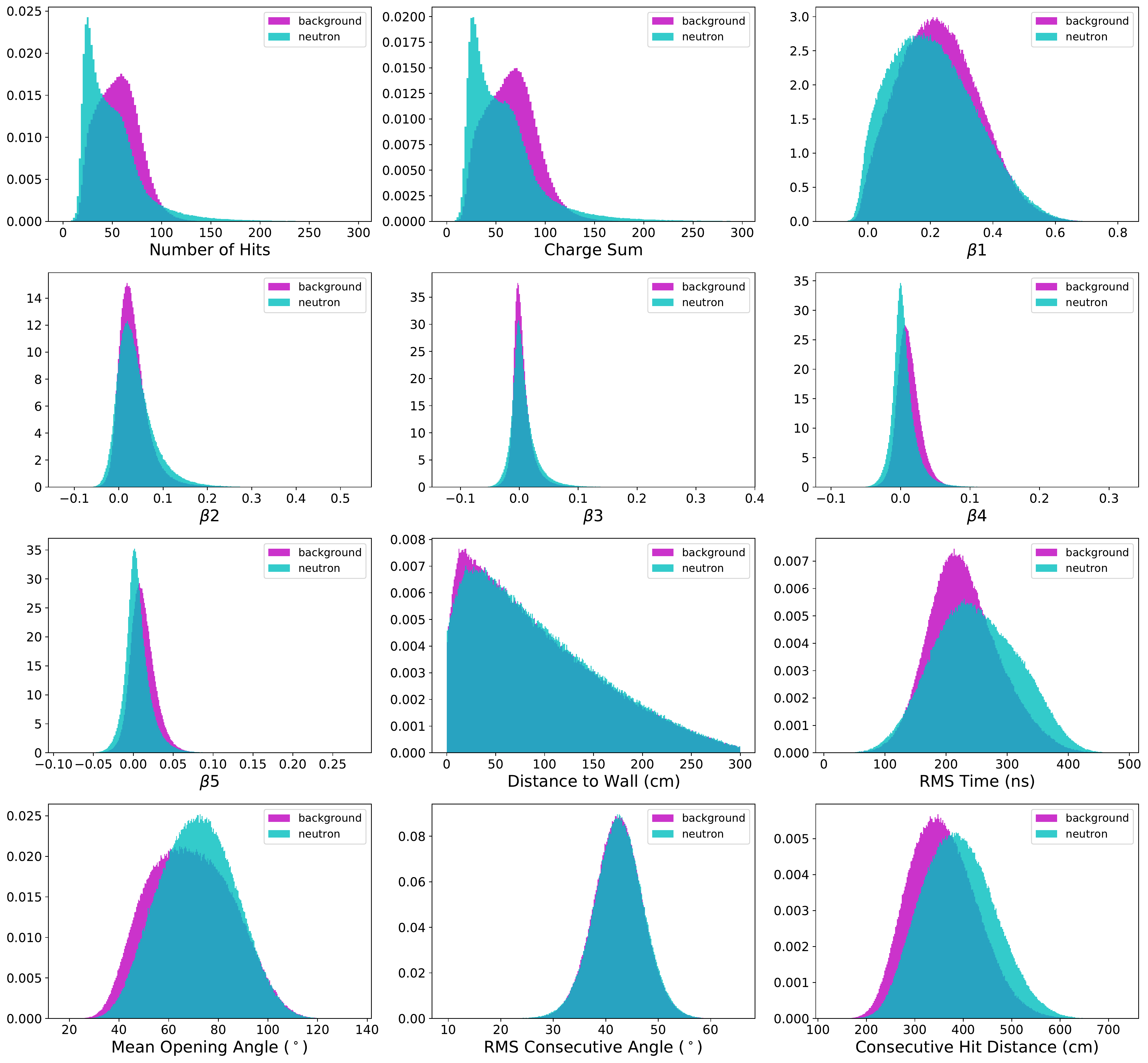} };
 \begin{scope}[x={(image.south east)},y={(image.north west)}]
    \node [anchor=north east] at (0.3,0.94) {(a)};
    \node [anchor=north east] at (0.63,0.94) {(b)};
    \node [anchor=north east] at (0.97,0.94) {(c)};
    \node [anchor=north east] at (0.3,0.69) {(d)};
    \node [anchor=north east] at (0.63,0.69) {(e)};
    \node [anchor=north east] at (0.97,0.69) {(f)};
    \node [anchor=north east] at (0.3,0.44) {(g)};
    \node [anchor=north east] at (0.63,0.44) {(h)};
    \node [anchor=north east] at (0.97,0.44) {(i)};
    \node [anchor=north east] at (0.3,0.19) {(j)};
    \node [anchor=north east] at (0.63,0.19) {(k)};
    \node [anchor=north east] at (0.97,0.19) {(l)};
     
 \end{scope}
\end{tikzpicture}  
  \caption{Comparison of twelve engineered features separated by neutron capture and spallation electron background events. The data consists of nearly 1.6 million events, generated by WCSim for the IWCD detector geometry.}
  \label{feature_diff_spallation}
\end{figure}

\subsection{Likelihood Baseline Analysis}
\label{likelihood section}

As shown in panels (a) and (b) of Fig.~\ref{feature_diff_spallation}, the difference in the total number of hits and charge sums between neutron and background electron events is the most obvious source of separability between these event types (the rest of the distributions in Fig.~\ref{feature_diff_spallation} will be discussed in the following sections). A statistical likelihood analysis based on these features was implemented to determine a baseline classification accuracy for later comparison against other machine learning approaches. The likelihood baseline classification accuracy was calculated by estimating the probability density function (PDF) of the neutron and electron events based on their hit and/or charge sum distributions and then classifying the events based on highest likelihood. The kernel density estimate (KDE) was used as an estimate of the underlying PDF for the corresponding distribution. The density of the KDE instance, once fit over a distribution of data, was then used to evaluate the event likelihood at a given point.

A univariate and multivariate approach were both applied, and the likelihood analysis was carried out depending on the evaluation type. For the `hits' evaluation type, the univariate KDE was calculated for neutron and electron events over the training set and then the events in the test set were classified based on highest probability between the neutron hits KDE and the electron hits KDE. For the `q\_sum' (charge sum) evaluation type, an identical process was undertaken, except the univariate KDEs were calculated for the neutron and electron events based on their charge sums over the training events. The final evaluation type, `q\_sum \& hits', involved calculation of multivariate KDEs for neutron and electron events on the training set for the combined 2-dimensional distribution of charge sums and number of hits combined. All events in the test set were then classified based on the highest density of the neutron and electron multivariate KDEs for every tuple of event hits and charge sums for that event type.

Table~\ref{likelihood classification table} shows the results of the likelihood classification approach using univariate and multivariate KDEs.  There is a high degree of correlation between the hits and q\_sum variables, and therefore not much difference is expected between these likelihood analyses.  Evaluation based on the univariate q\_sum KDE yielded the top accuracies for each dataset. The runtime cost of classifying events from highest KDE likelihood was approximately one hour and twenty minutes on average for the testing set, while fitting the univariate or bivariate KDEs to the training dataset only took a few minutes. 

\begin{table}[ht!]
\centering
\begin{tabular}{|c|c|c|}
\hline
 hits (1D) & q\_sum (1D) & q\_sum \& hits (2D) \\ \hline
 62.4         & 62.5        & 62.5                   \\ \hline
\end{tabular}
\caption{Classification accuracy of neutron capture and background electron events by comparison of the highest likelihood, as determined by the univariate hits KDE (hits), univariate charge sum KDE (q\_sum), and the multivariate KDE of charge sums and hits combined (q\_sum \& hits). KDE distributions were fit on the training set and applied on the test set.}
\label{likelihood classification table}
\end{table}


\section{Feature Engineering} 
\label{feature engy}

In machine learning, feature engineering is the process of applying domain knowledge to extract useful features from the original dataset. These features are often more useful than the raw data itself for predictive or analytic tasks. However, the features must be carefully selected  to extract as much information from the data as possible. Thus, a search was conducted for useful features in the domain of neutron capture in WC detectors. Relevant features were selected to aggregate information from each event, reducing the complexity of the dataset and extracting it into a more useful format. It was found that the classification performance of the XGBoost models significantly improved upon application to the aggregated features compared to the original dataset.

\subsection{Beta Parameters}

One way to quantify event topology is by the amount of anisotropy within the event with respect to the event vertex (the vertex position denotes the Cartesian coordinates of the start of the event). For comparison of neutron capture to background events, isotropy may be a discriminating factor due to the backgrounds being single electron events, while the neutron signal is multiple gammas from a neutron capture. Several isotropy parameters were considered for use in this study, including $\Theta_{ij}$, ``the average of the angles between each pair of PMT hits in an event with respect to the fitted vertex position," the correlation function ring inner product (CFRIP), which compares the angular correlation of the event to that of a perfect ring, and the beta parameters $\beta({\mathit{l}})$, defined similarly to $\Theta_{ij}$ but which make use of Legendre polynomials \citep{wilson}.

Both \citet{wilson} and \citet{dunmore} found the beta parameters to yield the most powerful discrimination based on event isotropy between different types of subatomic particle events. Following this result, the beta parameters were chosen as the measure of isotropy in this project. The definition for the \textit{l}-th beta parameter $\beta({\mathit{l}})$ is

\begin{equation}
    \beta({\mathit{l}}) = \langle P({\mathit{l}}) (\cos\theta_{ik}) \rangle _{i\neq k},
\end{equation}

\noindent where $\beta({\mathit{l}})$ is equal to the average of the \textit{l}-th Legendre polynomial $P({\mathit{l}})$ of the cosine of the angle $\theta_{ik}$ between every pair of hit PMTs in the event ($i \neq k$) with respect to the event vertex. For any of the beta parameters, a value of 0 indicates perfects isotropy, while higher absolute values indicate directionality and lower isotropy.  The beta parameter distributions for the datasets in this paper are shown in panels (c) through (g) of Fig.~\ref{feature_diff_spallation} for $\beta_1$ through $\beta_5$ respectively.  In practice, an event vertex would need to be calculated using an existing vertex reconstruction method. For the purpose of this study, the truth information is used for the exact event vertex position.

\subsection{Time of Flight}

The root-mean-square (RMS) time of flight was selected as an engineered feature to extract timing difference information from the data. The RMS time was calculated for a given event as the square root of the sum of the squared differences of every hit time from the average hit time per event, averaged over the number of hits for that event:

\begin{equation}
    t_{RMS}(x) =  \sqrt{\frac{\sum_{i=1}^{N(x)} (t_{i}(x) - t_{\mu(x)}) ^{2}}{N(x)}},
\end{equation}

\noindent where $i$ is an individual hit within the event $x$,
$N(x)$ is the number of hits in event $x$, $t_{i}$ is the recorded time of hit $i$, and $t_{\mu(x)}$ is the average hit time for the event.

The RMS time of flight, shown in panel (i) of Fig.~\ref{feature_diff_spallation}, has greater resistance to dark noise fluctuations (random hits before or after an event) and was found to show greater discrimination between signal and background compared to the overall time of flight.

\subsection{Distance to Wall}

The distribution of event vertex distance to the IWCD cylindrical tank wall, inspired by \citet{irvine}, was also explored as a potential discriminating feature between neutron capture and background events. For an underground WC detector such as Super-Kamiokande, which is located approximately one kilometer underground, a greater number of background events may originate at positions nearer the detector walls due to radiation from the surrounding rock. In the simulated IWCD data, there is a slightly greater occurrence of neutron capture events in the region of 50~cm to 300~cm from the tank wall, as seen in panel (h) of Fig.~\ref{feature_diff_spallation}.

\subsection{Mean Opening Angle}
\label{moa section}

The Cherenkov emission from relativistic photons in water is emitted on a cone with respect to the origin of radiation. The angle of emission is dependent on the kinematic properties of the incident charged particles. The mean opening angle from the event vertex varies on average for different types of particle interactions, making this metric another possible discriminant to improve neutron tagging performance. Following the definition in \citet{irvine}, this mean opening angle is calculated as the mean value of the angles between every hit PMT vector and the true vertex position within the given event:

\begin{equation}
    \Theta_{\mu}(x) = \frac{\sum_{i=1}^{N(x)} \Theta(\vec{p_{i}}, \vec{p_{0}})} {N(x)} = \frac{\sum_{i=1}^{N(x)} \arccos{(\frac{\vec{p_{i}} \cdot {\vec{p_{0}}}}{|\vec{p_{i}}| \cdot |\vec{p_{0}}|})}}{{N(x)}},
\end{equation}

\noindent where $\Theta_{\mu}(x)$ is the mean opening angle for the event $x$, $N(x)$ is the number of hits, $\vec{p_{i}}$ is the (x, y, z) position of the $i$-th hit, $\vec{p_{0}}$ is the (x, y, z) position of the event vertex and $\Theta(\vec{p_{i}}, \vec{p_{0}})$ is the angle between $\vec{p_{i}}$ and $\vec{p_{0}}$, computed as the quotient of the dot product by the product of their magnitudes. 

The mean opening angle metric is largely influenced by the event energy. Discrimination is observed between the distributions seen in panel (j) of Fig.~\ref{feature_diff_spallation} due to a combination between the event energy and topological distribution of the hits throughout the event. The electron events in the background dataset have lower energies, but the hits are more sparsely distributed.  Evidently, the net effect is that the neutrons end up with a higher peak mean opening angle than the background events in the dataset, on average.

\subsection{Consecutive Hit Angular RMS}

Another potential neutron tagging feature discriminant, again inspired by \citet{irvine}, is the root-mean-squared consecutive angle of an event. True background hits, for example from radioactive background sources, often contain spatially compact clusters of hits. On the other hand, Cherenkov photons from neutron capture events would be expected to propagate more uniformly within the average opening angle of the radiation emission cone. The RMS difference of angle between temporally consecutive hits can extract information on these angular differences between event types. The RMS angle is calculated by first sorting all PMT hits chronologically within a given event, then computing the sum of the squared differences of the angles between consecutive events from the mean consecutive angular difference, averaged over the number of hits for the event and square rooted, as

\begin{equation}
    \Theta_{RMS}(x) = \sqrt{\frac{\sum_{i=1}^{N(x)-1} (\Theta(\vec{p_{i}}, \vec{p_{i+1}}) - \Theta_{\mu})^{2}} {N(x)}} = \sqrt{\frac{\sum_{i=1}^{N(x)-1} (\arccos{(\frac{\vec{p_{i}} \cdot {\vec{p_{i+1}}}}{|\vec{p_{i}}| \cdot ||\vec{p_{i+1}}|}}) - \Theta_{\mu}) ^{2}} {N(x)}}, 
    \label{rms angle}
\end{equation}

\noindent where $\Theta_{RMS}(x)$ is the RMS consecutive angle for the event $x$, $N(x)$ is the number of hits, $\vec{p_{i}}$ is the (x, y, z) position of the $i$-th hit, $\vec{p_{i+1}}$ is the (x, y, z) position of next consecutive hit in time order {i+1}, $\Theta_{\mu}$ is the average angle between consecutive hits in the event and $\Theta(\vec{p_{i}}, \vec{p_{i+1}})$ is the angle between $\vec{p_{i}}$ and $\vec{p_{i+1}}$.

For events with more scattering, clustering and reflections, the distributions of RMS consecutive angles will be higher on average, and vice versa. Panel (k) shows that there is little difference between neutron and background signals, which is expected since our simulation uses a uniform distribution of background events.  For a background source more inclusive of clustering, the discrimination extent is expected to be greater for the RMS angular metric.

\subsection{Consecutive Hit Distance}

In studying the event displays of the neutron capture and background events, it was observed that the positional distributions of hits tended to be more widespread in neutron capture events. Given two events of different type with similar numbers of hits, the neutron capture event could be reasonably well differentiated by eye by selecting the event with greater average distance between hits. To compute the average consecutive hit distance, the hits within a given event were first sorted chronologically in time, then the Euclidean distances between consecutive hits were summed and averaged over the number of hits within the event as

\begin{equation}
\begin{multlined}
    hd_{\mu}(x) = \frac{\sum_{i=1}^{N(x)-1} dist(\vec{p_{i}}, \vec{p_{i+1}})} {N(x)} = \\
    \frac{\sum_{i=1}^{N(x)-1} \sqrt{({p_{x(i)}} - {p_{x(i+1)}})^{2} + ({p_{y(i)}} - {p_{y(i+1)}})^{2} + ({p_{z(i)}} - {p_{z(i+1)}})^{2}}} {N(x)},
    \label{average hit dist}
\end{multlined}
\end{equation}

\noindent where $hd_{\mu}(x)$ is the average consecutive hit distance for the event $x$, $N(x)$ is the number of hits, $\vec{p_{i}}$ is the (x, y, z) position of the $i$-th hit, $\vec{p_{i+1}}$ is the (x, y, z) position of the next consecutive hit in time order {i+1} and $dist(\vec{p_{i}}, \vec{p_{i+1}})$ is the Euclidean distance between consecutive hits.

The difference of consecutive hit distance was a good discriminator, as seen in panel (l) of Fig.~\ref{feature_diff_spallation}.  This difference in consecutive hit distance is likely due to the differing nature of the particle interactions, in which the cascade of gammas from the neutron capture leads to greater spatial separation of hits throughout the detector, on average, when compared to the electron background hits.

\subsection{XGBoost Results}
\label{xgb results}

The XGBoost gradient boosting decision tree model was applied to the task of learning from the features engineered in Section \ref{feature engy}.A grid search was applied to tune the model hyperparameters, including the maximum tree depth \textit{max{\_}depth}, the minimum tree node weight \textit{min{\_}child{\_}weight}, the training data subsampling ratio \textit{subsample}, the tree column subsampling ratio \textit{colsample{\_}bytree} and the learning rate \textit{eta}. The grid search sequentially iterated over  relating parameters pairs and applied four-fold cross-validation to improve outcome reliability. The relating pairs were \textit{max{\_}depth} and \textit{min{\_}child{\_}weight}, followed by \textit{subsample} and \textit{colsample{\_}bytree}. The learning rate was adjusted independently. For each hyperparameter combination, XGBoost's native cross-validation function was used to train the model over a maximum of 1250 boosting rounds, and early stopping was used to cancel model training if performance did not improve over twenty consecutive rounds. 

\begin{figure}[!htb]
  \begin{center}
    \includegraphics[width=0.55\textwidth]{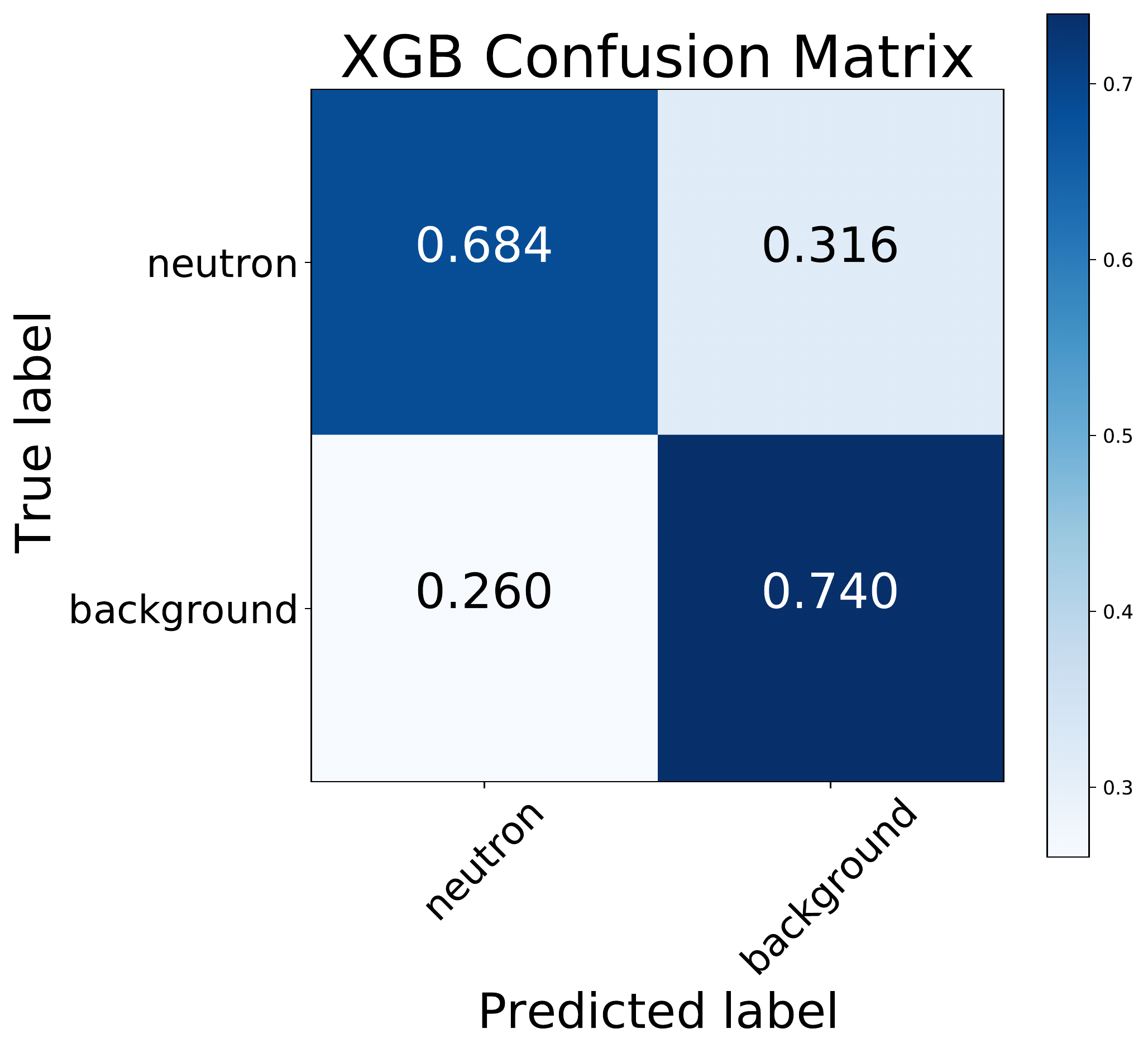}
  \end{center}
  \caption{Confusion matrix for the XGBoost model trained on the dataset of neutron capture and spallation background electron events.}
  \label{cm_XGBoost_spallation}
\end{figure}

Applying this technique, the optimal tree complexity was found with \textit{max{\_}depth} of 11 and a \textit{min{\_}child{\_}weight} of 1, the optimal sampling ratios were found with a \textit{subsample} ratio of 0.7 and a \textit{colsample{\_}bytree} ratio of 1.0, and the learning rate was tuned to 0.007. The optimized XGBoost model was then trained on a consistent 80\% training dataset, optimized against an indepedent 10\% validation dataset and tested against a 10\% holdout test set. The model obtained train, validation and test accuracies of 73.0\%, 71.5\% and 71.4\% respectively, and an ROC AUC score of 0.784.  Although the training accuracies are generally slightly higher than the test accuracy, the extent of overfitting was not too severe and may be decreased by using a smaller number for early stopping. The XGBoost model construction for any of the 80\% training sets was found to take approximately 45 to 60 minutes, depending on the number of trees constructed before early stopping. Figure \ref{cm_XGBoost_spallation} displays the confusion matrix, which shows that the true positive rate (neutron sensitivity) is significantly lower than the true negative rate (neutron specificity). 

\begin{figure}[!htb]
  \textbf{}
  \centering
  \includegraphics[width=0.85\linewidth]{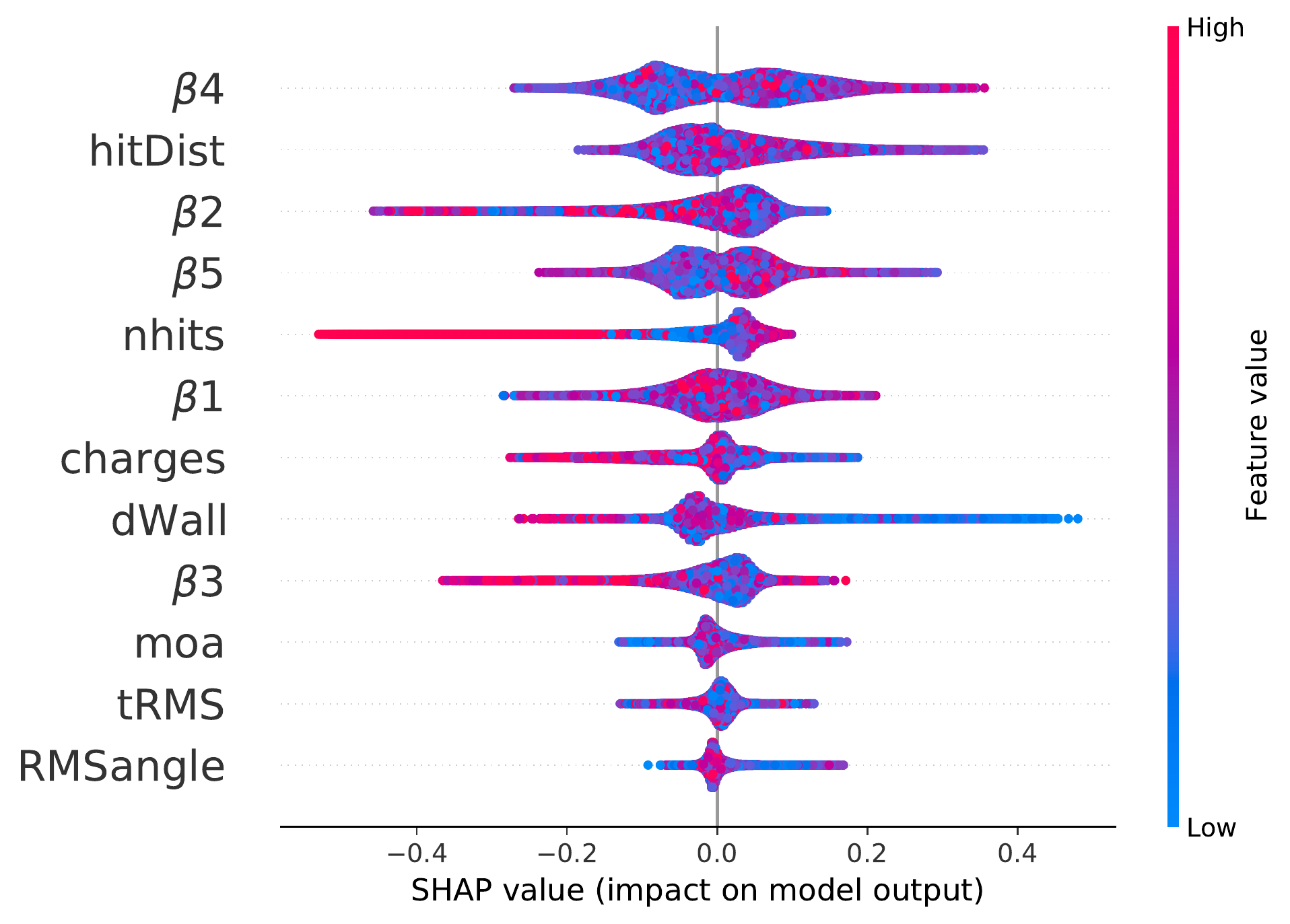}
  \caption[SHAP beeswarm plot for the neutron capture and spallation background dataset.]{Beeswarm plot of SHAP values for the neutron capture and spallation background dataset, simulated using WCSim for the IWCD tank geometry. The SHAP value for each feature in every event is plotted as a dot in the plot, where the x-axis position corresponds to the SHAP value and the colorbar shows the feature value (blue is low, red is high). High SHAP values influence the model output towards 1 (electron-like event) and low SHAP values (negative) influence the model outputs toward 0 (neutron-like event).}
  \label{beeswarm_spallation}
\end{figure}

SHAP was used to understand the relative importances of the dozen features contributing to the XGBoost model. The SHAP values are applicable both locally, to a single event, and globally, to a conglomerate of events. While various visualizations using SHAP are possible, the `beeswarm' plot in particular is useful in showing the range and density of SHAP values for individual features. Figure \ref{beeswarm_spallation} shows the beeswarm plot over all events in the neutron capture and spallation electron background dataset. For this plot, the SHAP value for each feature in every event is plotted as a single dot. Bulges in a row indicate areas of larger density. Higher SHAP values influence the model output towards 1 (electron-like event) and low SHAP values (negative) influence the model outputs toward 0 (neutron-like event). The features are arranged on the vertical axis by feature importance, with the most important features (by average absolute SHAP value) on the top and the least important features on the bottom. Each feature value is plotted with a colour corresponding to its position within its numeric range.

Several distinctive patterns from Fig. \ref{beeswarm_spallation} are discernible. For a high number of hits, the SHAP value is uniformly negative. Correspondingly, in panel (a) of Fig. \ref{feature_diff_spallation}, it is clear that events with more than approximately 100 hits are uniformly neutron events (top-left plot). For the wall distance, it is clear from panel (h) of Fig. \ref{feature_diff_spallation} that there are is a slight over-representation of background events at distances close to the wall. The XGBoost model clearly notices this difference, as events with lower wall distances mostly have higher SHAP values, meaning the model output value is pushed higher to 1 (electron-like event). 

The beeswarm Fig. \ref{beeswarm_spallation} also reveals a notable difference between the lower order ($\beta$1, $\beta$2, $\beta$3) and higher-order ($\beta$4, $\beta$5) isotropy parameters. $\beta$1, $\beta$2 and $\beta$3 both have single mode representations in the beeswarm plot, in which there is a single bulge. Higher values for these parameters also attribute the output toward a neutron classification, on average. This correspondence may be seen by the feature differences of panels (c) to (g) of Fig. \ref{feature_diff_spallation}. Alternately, $\beta$4 and $\beta$5 have two main modes (bulges) in the SHAP value beeswarm plot, indicating two main regions with SHAP values of a similar range. For $\beta$5, a clear distinction is seen between lower values of $\beta$5, attributed toward neutron events, and higher values of $\beta$5, attributed toward electron events. While this difference is clear, the SHAP values themselves are lower, showing a smaller output impact. This small difference is observable in panel (g) of Fig. \ref{feature_diff_spallation} for $\beta$5. 

The $\beta$4 parameter has a similar double-moded pattern in the beeswarm plot, but the attributed difference is smaller for lower and higher values of the parameter. However, $\beta$4 still has the greatest average absolute SHAP value, and therefore the greatest average impact on the model output. In general, $\beta$4, mean consecutive hit distance, $\beta$2, $\beta$5 and number of hits respectively were the top five most important features in determining event outcomes.


\section{Graph Neural Network Application}

In this study, the PyTorch Geometric (PyG) library was used to apply graph neural network models to the IWCD dataset \citep{pytorchgeometric}. This particular library was chosen for its ease of use, breadth of graph network models available, data loading tools and GPU support. During training, at regular intervals, the model was applied to the validation dataset to check for under fitting or overfitting. After the model was trained, it was applied on the test dataset and evaluation metrics were computed. Model parameters were updated using Adam optimization \citep{adam} with cross-entropy loss. Training was carried out on a Quadro P2000 GPU. 


\subsection{Graph Convolutional Network (GCN)}

For a neutron capture or background event, the hit PMTs may be represented as graph nodes, with each node containing the features of hit time, deposited charge and the 3-dimensional position and orientation of the hit PMT. Since the number of hits varies for every event, the graphs could either vary in size (non-padded graph) or zero padding could be added. Graph padding, along with edge weighting and node connectivity, were three hyperparameters of graph construction investigated in this research. Within the GCN framework, model performance was compared against padded versus non-padded graphs, edge weighted (inversely proportional to distance) versus uniform weights, and the fully connected versus $k$ nearest neighbour graph.

To begin, the GCN model was tested on graphs constructed using a padded, fully connected (every node connected to every other node) representation with all edge weightings set to a value of one. This setting was used to adjust parameters of the GCN architecture, leading to the configuration of two alternating layers of GCN convolutional filtering and activation computation, with 24 and 8 compute nodes in the first and second hidden layers respectively. This was followed by max pooling and the log softmax output from a fully connected layer with two neurons in the output layer. GCN results were obtained by training with a batch size of 32, learning rate of 0.0003 and learning rate decay of 0.001\%.

The first graph construction comparison tested whether the GCN model learned better on padded graphs or variable-size graphs without padding. The results of training the GCN model for \textit{padded} and \textit{non-padded}, fully connected graphs with uniform edge weightings are shown in the first two rows of Table \ref{gcn1}. The performance was higher with for padded graphs, with an average test accuracy improvement of 3.2\%.

\begin{table}[!htb]
\centering
\begin{tabular}{|c|c|c|c|c|}
\hline
Type& Train Accuracy & Validation Accuracy & Test Accuracy & ROC AUC \\ \hline
Padded                                      & 61.3                                   & 63.1                                     & 63.1                               & 0.667                             \\ \hline
Non-padded                                      & 58.5                                   & 59.8                                     & 59.9                               & 0.628                             \\ \hline
Padded ($1/d^2$) & 59.7                                   & 61.3                                     & 61.4                               & 0.632  \\ \hline
\end{tabular}
\caption{GCN model applied to padded and non-padded, fully connected,  uniformly edge weighted, and inverse square distance ($1/d^2$) weighted graphs for the simulated IWCD neutron capture and background datasets.}
\label{gcn1}
\end{table}

While accuracy was higher for the padded graphs, the runtime was also considerably longer due to the significantly greater number of connections and message passing operations in the padded graphs. Per epoch, the padded graphs took 6 hours to train, while the non-padded graphs took only 14 minutes. The non-padded graph GCN model was trained over 75 epochs and approximately 17 hours, while the padded GCN model was trained over 5 epochs and approximately 30 hours. A higher number of epochs was not found to improve the performance for either model. A summary of runtimes is presented in table \ref{training times}.

Next, edge weighting was tested for the GCN model to see if edge values related to physical distance could provide a learning advantage over uniform edge weightings set to a tensor of ones. The results of training the GCN on fully connected, padded, inverse-distance edge weighted graphs is shown in the last row of Table \ref{gcn1}. Runtimes are nearly identical to the same model with fixed edge weights. With distance weighted edges, the test accuracy was 1.7\% lower than the corresponding result for graphs with uniform edge weights. The edge weightings possibly overcomplicate the GCN model on the scale of the $\sim 10^5$ node connections for a given event.

Overall, the GCN model was found to perform best on static, fully connected, uniform edge weighted graphs. The results are shown in Table \ref{gcn1}. This GCN configuration has comparable metrics to the highest likelihood baseline (see Table \ref{likelihood classification table}, q\_sum), with 0.6\% higher accuracy on the spallation set. Therefore, it may be concluded that the GCN model was largely learning to classify events based on the trivial number of hits, and that it failed to significantly learn from the geometric differences of neutron capture to electron background hit patterns. Adding any additional network layers to the GCN models was also found to worsen performance, presumably as the extra filtering step oversmoothes the node representations. Overall, the GCN model is perhaps better suited to contexts with fewer and more meaningful node connections and edge weightings, such as social networks and credit fraud identification, rather than the `point cloud' representation of the particle events.

\subsection{Dynamic Graph Convolutional Neural Network}

\begin{figure}[!htb]
  \begin{center}
    \includegraphics[width=0.97\textwidth]{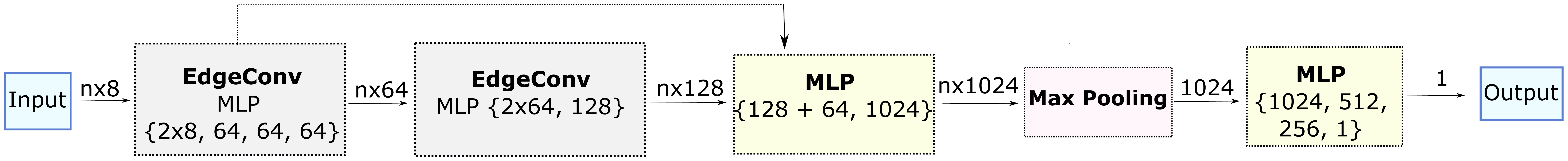}
  \end{center}
  \caption{Applied DGCNN architecture for neutron capture and electron background event discrimination. Two dynamic edge convolutional blocks were applied, followed by a fully connected layer, global max pooling, and a final multi-layer perceptron layer.}
  \label{fig:dgcnn_diagram}
\end{figure}

The DGCNN model was the next graph network model applied to the particle classification task. Described in Section \ref{dgcnn section}, the DGCNN model was selected for its ability to learn from point cloud data specifically. The network architecture configuration was set to the default from the PyTorch Geometric example documentation, which consisted of the following: two dynamic edge convolution layers followed by a fully connected layer, a global max pooling layer and a final MLP to yield the output class probabilities. The first edge convolution layer applied an MLP on input node features with three layers of 64 compute units each. The second edge convolution layer took the output of the first as input and applied an MLP with 128 activation units. In both cases, the MLP was applied to every node pair ($n \ast 2$ pairs for $n$ nodes in an event) over the $k$-nn ($k$ nearest neighbour) graph representation of each node, and the representations were updated by pooling the learned edge features. After the EdgeConv blocks, a fully connected layer concatenated the 64 and 128 unit features from the first two dynamic edge convolution layers and yielded 1024 activations. Global max pooling was applied over the $n$ nodes to reduce the representation from $n \ast 1024$ to only $1024$. The final MLP then passed this information into final layers of 512, 256 and 2 activation nodes respectively and the softmax of the output was applied to calculate the output the binary classification probabilities. Note that for every fully connected layer throughout the network, the activations were calculated using the ReLU activation function and batch normalization \citep{batchnorm} to reduce overfitting. The model description is represented by Fig.~\ref{fig:dgcnn_diagram}.

With the fixed architecture as described above, the number of nearest neighbours $k$ in the DGCNN dynamic edge convolution blocks were adjusted over multiple runs to compare performance. Table \ref{dgcnn1} shows the results of applying the DGCNN model on the spallation background dataset with the $k$ hyperparameter varying from 10 to 30 in increments of 5. The resulting accuracies were largely the same for $k = 15$ to $k = 30$, while $k = 10$ neighbours led to a slightly lower accuracy. Among the range of $k = 15$ to $k = 30$, $k = 25$ yielded the highest ROC AUC score of 0.797, although the 0.001 difference compared to the other $k$ values in the range was not necessarily statistically significant. 

Regarding the statistical sensitivity, for the 80,000 validation events in each sample the statistical uncertainty is 0.4\%. Since the same datasets are being reused in changing the $k$ values, there is some cancellation of uncertainty due to correlation through using the exact same data.  A conservative estimate of 0.4\% uncertainty could be used here, since relying on correlation by not having generated enough data to compare the methods using independent data may mean fitting to some peculiarity of the dataset, rather than a generally useful difference.

\begin{table}[!htb]
\centering 
\begin{tabular}{|c|c|c|c|c|}
\hline
k neighbours & Train Accuracy & Validation Accuracy & Test Accuracy & ROC AUC \\ \hline
10           & 70.9           & 72.0                & 71.9          & 0.792   \\ \hline
15           & 71.8           & 72.2                & 72.3          & 0.796   \\ \hline
20           & 71.7           & 72.3                & 72.3          & 0.796   \\ \hline
25           & 71.8           & 72.4                & 72.4          & 0.797   \\ \hline
30           & 71.4           & 72.4                & 72.4          & 0.796   \\ \hline
\end{tabular}
\caption{DGCNN model classification accuracies for variations of the number of nearest neighbours $k$ in the DGCNN dynamic edge convolution blocks from 10 to 30 in increments of 5.}
\label{dgcnn1}
\end{table}

While there was minimal performance difference for the range of $k = 15$ to $k = 30$, there was however a difference in the training times, as shown in table \ref{training times}. This was expected as, for $n$ $f$-dimensional input nodes, an $n \ast k \ast a_{n}$ -dimensional tensor is generated before pooling across the neighbouring edge features for every dynamic edge convolution block. Therefore, the total number of training parameters increases significantly for every increment of nearest neighbours $k$. There was a sharp increase in training time after about $k = 20$ and more than a doubling in overall training time from $k = 10$ to $k = 30$.

Given the results in Table \ref{dgcnn1} and the processing times required, $k = 15$ is a good compromise between training time and classification accuracy. However, when training time is not a significant impediment, $k = 25$ might be used to optimize results.

\section{Discussion}

For all models, consistent training, validation and test datasets were constructed in an 80\%, 10\% and 10\% ratio. Models were optimized against the validation data and metrics were reported for the holdout test dataset, ensuring that differences in model performance were not due to random distributions of the data. Compared to the likelihood statistical baseline, the DGCNN model results in Table \ref{dgcnn1} showed an accuracy improvement of 9.9\%.  The $\sim10$\% classification accuracy improvement strongly indicates the capability of the DGCNN model to learn from event topology and other, more subtle factors than the number of hits and overall sum of charges within the event. The dynamic method of graph construction with the DGCNN model, which shuffles the groupings of every node with its other nearest neighbour nodes in semantic space, allows the diffusion of nonlocal information throughout the graph. This ostensibly allows the DGCNN model to learn global event topology in a way which the GCN model, restricted to operating over fixed input graphs, was not able to. 

Overall, the DGCNN also slightly outperformed the best XGBoost model, representing an improvement in accuracy of 0.7\% and ROC AUC score of 0.007. The test accuracy results for all approaches undertaken in this study, including the likelihood baseline analysis, XGBoost with feature engineering and the GCN and DGCNN models are presented in Table \ref{Tab:overall summary table}. The best accuracy for neutron versus background separation was 72.4\% using DGCNN.

\begin{table}[!htb]
\centering
\begin{tabular}{|c|c|c|c|c|}
\hline
Dataset Background Source & Likelihood & XGBoost       & GCN  & DGCNN         \\ \hline
Spallation                & 62.5       & 71.4          & 63.1 & \textbf{72.4} \\ \hline
\end{tabular}
\caption{Overall accuracies for neutron capture versus electron background classification for the likelihood analysis (Likelihood), XGBoost, GCN and DGCNN methods.}
\label{Tab:overall summary table}
\end{table}

\begin{figure}[!htb]
  \begin{center}
    \includegraphics[width=0.55\textwidth]{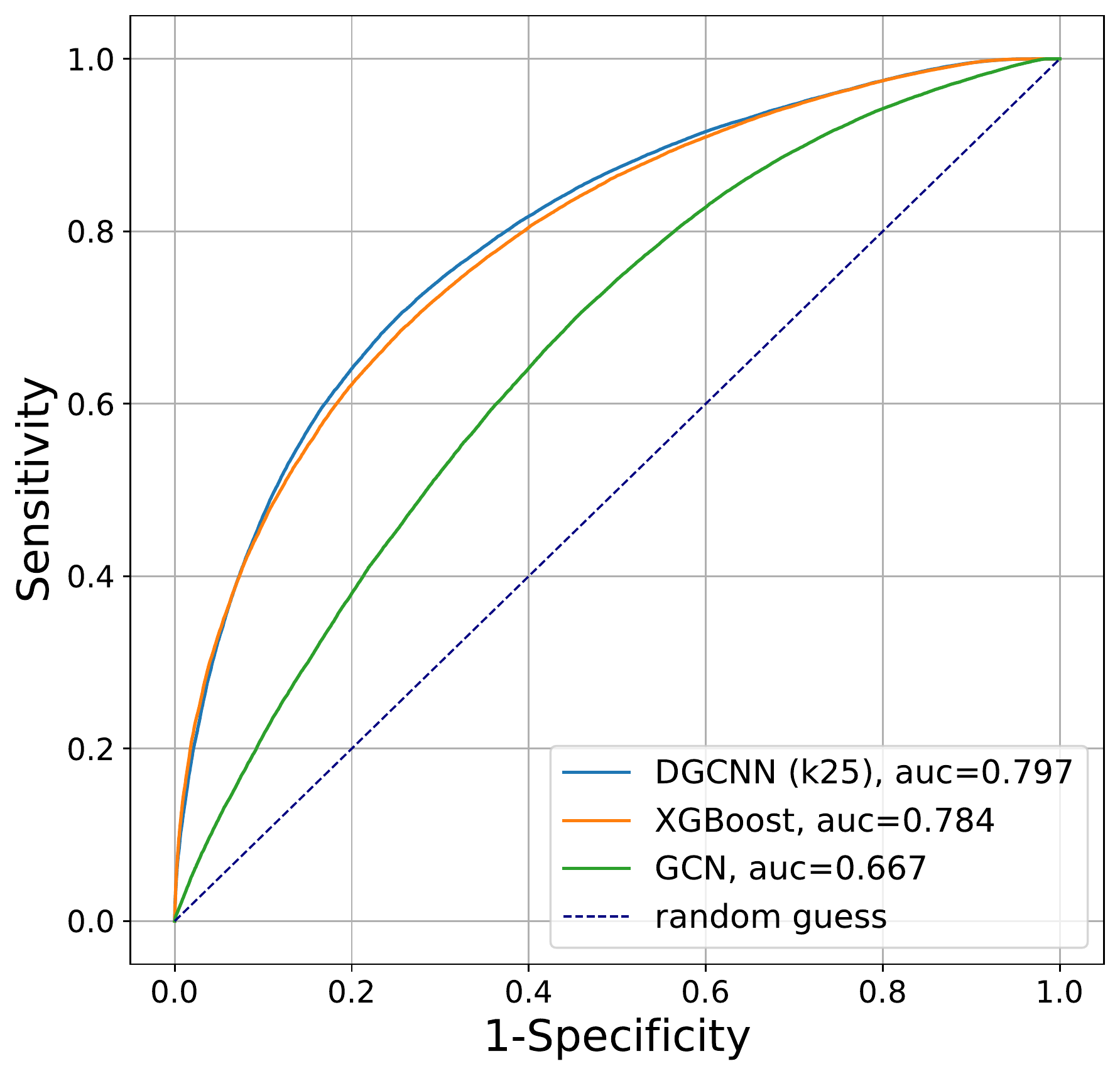}
  \end{center}
  \caption{Comparison of the ROC curves for the XGBoost, GCN and DGCNN results presented in this paper.}
  \label{fig:roc}
\end{figure}

The receiver operating characteristic (ROC) curves, which plots the true positive rate (sensitivity) against the false positive rate (1 - specificity) for a binary classification problem, are shown in Fig.~\ref{fig:roc} for the different machine learning methods studied in this paper. The ROC AUC (area under the curve) from XGBoost is 0.784, from GCN is 0.667, and from DGCNN with k=25 is 0.797, showing that consistent with the accuracies presented earlier, the DGCNN had the best performance.

\begin{table}[ht!]
\centering
\begin{tabular}{|c|c|c|c|}
\hline
Model            & Epochs & Total Runtime (min) & Time Per Epoch (min) \\ \hline
XGBoost          & 1450   & 50                  & 0.036                \\ \hline
DGCNN (k=10)     & 25     & 1980                & 1.32                 \\ \hline
DGCNN (k=15)     & 25     & 2100                & 1.4                  \\ \hline
DGCNN (k=20)     & 25     & 2700                & 1.8                  \\ \hline
DGCNN (k=25)     & 25     & 3420                & 2.28                 \\ \hline
DGCNN (k=30)     & 25     & 3960                & 2.64                 \\ \hline
GCN (non-padded) & 75     & 1020                & 13.6                 \\ \hline
Likelihood Ratio & 1      & 80                  & 80                   \\ \hline
GCN (padded)     & 5      & 1800                & 360                  \\ \hline
\end{tabular}
\caption{Comparison of training times for the different models applied in this study, sorted in ascending order by training time per epoch.}
\label{training times}
\end{table}

\section{Conclusions}

This paper has presented a search to improve the classification performance of neutron capture versus background identification in WC detectors using techniques in machine learning.  To provide a performance baseline, a statistical model was applied to classify events using maximum likelihood of kernel density estimates of the main event type discriminants, namely the number of hits and charge sums. The  baseline accuracy was found to be 62.5\%.


Next, a series of features were engineered from the datasets. Besides number of hits and charge sums, the beta parameters $\beta$1-$\beta$5 were created to capture event isotropy. The mean opening angle, event vertex distance to wall, RMS consecutive hit angle and mean consecutive hit distance were also computed to summarize event topology and the RMS event time was added to capture timing discrimination. Gradient boosted decision trees were applied on these engineered features using the XGBoost algorithm. The XGBoost model hyperparameters were tuned using grid search, yielding a test accuracy of 71.4\% which represented an 8.9\% improvement over the baseline approach. SHAP analysis of these model outputs revealed useful information. The $\beta$2, $\beta$4, $\beta$5, number of hits and consecutive hit distance parameters were consistently rated most important, as measured by the mean absolute SHAP value. 

Drawbacks to the XGBoost and the feature engineering approach include preprocessing time to calculate the feature values and the fact that the calculation of several features relies on the event vertex position. For this research, the true vertex position was taken from the simulation information, but in reality a vertex reconstruction algorithm would need to be used, introducing some uncertainties into the equations. As an alternative approach, deep learning was implemented via the GCN and DGCNN graph neural network models.


The GCN model was tested with a variety of graph construction approaches, including static versus non-static graphs, uniform edge weighting versus scaled edge weights, and fully connected versus partially connected graphs. Of all these cases, the best test accuracy obtained was 63.1\% using the fully connected, zero padded, uniform edge weighted graphs, which was nearly identical to the baseline likelihood accuracy. This indicated that the GCN was likely learning mainly from the number of hits in the event, or the number of non-zero nodes.

The DGCNN model, however, was found to have significantly improved neutron tagging performance above the baseline accuracy. The DGCNN number of neighbours hyperparameter $k$ was tuned, and the reported accuracy was found to be 72.4\%, representing an improvement of 9.9\% over the likelihood analysis. Thus, DGCNN slightly outperformed XGBoost on the classification of neutron versus background. DGCNN also retains the advantage of not requiring any preprocessing or prior knowledge. On the other hand, XGBoost provides a much greater level of model interpretability. Furthermore, once the engineered features have been computed, the training time of XGBoost for the datasets used in this study was within the range of only 45 minutes to one hour, much faster than the DGCNN model which took from 30 to over 60 hours, depending on the value of $k$. However, DGCNN was trained over only a single GPU, and using multiple GPUs could reduce the runtime significantly. Table \ref{Tab:overall summary table} shows the overall results of XGBoost, GCN and DGCNN compared to the likelihood baseline.

Overall, both XGBoost with feature engineering and DGCNN show promise in improving neutron tagging efficiency in WC detectors. In particular, the application of these methods in the IWCD might help reduce systematic uncertainties for the Hyper-Kamiokande detector, which it turn could advance our understanding of neutrino physics and the Standard Model itself. In future, the network architecture of the DGCNN model could be further optimized. 

For practical purposes, given that these models were developed for data simulation, another reasonable next step would include the deployment of these models for neutron tagging in active WC detectors. This would test if the models are transferable for real use cases. Also, these models could be incorporated into a pipeline which tests for the coincidence of neutron capture and positron rings within a timescale indicative of neutrino inverse beta decay. While the development of improved neutron tagging is desirable, the ultimate goal is to trace back to the originating neutrino to probe deeper into the unknowns of neutrino physics. An end-to-end network could thus be deployed using the neutron tagging models developed in this research to better identify the neutrinos themselves in the overarching process of the neutrino inverse beta decay.

\section*{Conflict of Interest Statement}

The authors declare that the research was conducted in the absence of any commercial or financial relationships that could be construed as a potential conflict of interest.

\section*{Author Contributions}

This paper presents the research conducted by Matt Stubbs, whose thesis this paper is based on~\cite{matt}.  The other authors in this paper contributed to the development of the research at weekly meetings.  The initial implementation of the GNN code was prepared by John Walker, and datasets were prepared by Nick Prouse.  

\section*{Funding}
The funding for this research is from the Canadian National Science and Engineering Council (NSERC).  Production of the simulation datasets was done with the support of Compute Canada resources.

\section*{Acknowledgments}
We acknowledge the WatChMaL, Super-Kamiokande, and Hyper-Kamiokande collaborations, on whose shoulders we stand in coming up with the idea for this study, and with whom many of the authors are collaborators. This research was enabled in part by support provided by Cedar Compute Cluster (\url{https://docs.alliancecan.ca/wiki/Cedar}) and the Digital Research Alliance of Canada (\url{alliancecan.ca}).

\section*{Data Availability Statement}
The simulated datasets used for this study are part of ongoing research by the WatChMaL collaboration and are not publicly available.

\bibliographystyle{frontiersinHLTHFPHYS} 
\bibliography{test}


\end{document}


\onecolumn
\firstpage{1}

\title[Supplementary Material]{{\helveticaitalic{Supplementary Material}}}

\maketitle

\section{Supplementary Data}

Supplementary Material should be uploaded separately on submission. Please include any supplementary data, figures and/or tables. All supplementary files are deposited to FigShare for permanent storage and receive a DOI.

Supplementary material is not typeset so please ensure that all information is clearly presented, the appropriate caption is included in the file and not in the manuscript, and that the style conforms to the rest of the article. To avoid discrepancies between the published article and the supplementary material, please do not add the title, author list, affiliations or correspondence in the supplementary files.

\section{Supplementary Tables and Figures}

For more information on Supplementary Material and for details on the different file types accepted, please see \href{http://home.frontiersin.org/about/author-guidelines#SupplementaryMaterial}{the Supplementary Material section} of the Author Guidelines.

Figures, tables, and images will be published under a Creative Commons CC-BY licence and permission must be obtained for use of copyrighted material from other sources (including re-published/adapted/modified/partial figures and images from the internet). It is the responsibility of the authors to acquire the licenses, to follow any citation instructions requested by third-party rights holders, and cover any supplementary charges.


\subsection{Figures}


\begin{figure}[htbp]
\begin{center}
\end{center}
\caption{ Enter the caption for your figure here.  Repeat as  necessary for each of your figures}\label{fig:1}
\end{figure}

\begin{figure}[htbp]
\begin{center}
\end{center}
\caption{This is a figure with sub figures, \textbf{(A)} is one logo, \textbf{(B)} is a different logo.}\label{fig:2}
\end{figure}

